\documentclass[copyright,creativecommons]{eptcs}
\providecommand{\event}{ICE'09} 
\providecommand{\volume}{??}
\providecommand{\anno}{2009}
\providecommand{\firstpage}{1}
\providecommand{\eid}{??}

\usepackage{mathpartir}
\usepackage{url}
\usepackage{amssymb}
\usepackage{amsthm}
\usepackage{mathbbol}
\usepackage{stmaryrd}
\usepackage{empheq}
\usepackage{breakurl}

\newif\ifproofs
\proofstrue
\proofsfalse

\newif\ifcomments
\commentstrue
\commentsfalse

\newcommand{\eoe}{
  \renewcommand{\qedsymbol}{\hbox{$\blacksquare$}}
  \qed
}

\newcommand{\ccs}{\textsc{ccs}}

\newcommand{\contract}{\eta}
\newcommand{\varcontract}{\theta}
\newcommand{\client}{\rho}

\newcommand{\process}{P}
\newcommand{\varprocess}{Q}
\newcommand{\clientp}{R}
\newcommand{\type}{t}
\newcommand{\vartype}{s}
\newcommand{\val}{\mathsf{v}}
\newcommand{\expr}{e}

\newcommand{\tempty}{\mathbb{0}}
\newcommand{\tbasic}{\mathtt{B}}
\newcommand{\tint}{\mathtt{Int}}
\newcommand{\treal}{\mathtt{Real}}
\newcommand{\tbool}{\mathtt{Bool}}
\newcommand{\tstring}{\mathtt{String}}
\newcommand{\tdate}{\mathtt{Date}}
\newcommand{\taddress}{\mathtt{Address}}

\newcommand{\ttrue}{\mathtt{true}}

\newcommand{\cwin}{\mathbf{1}}
\newcommand{\cnull}{\mathbf{0}}

\newcommand{\ain}[1]{{?#1}}
\newcommand{\aout}[1]{{!#1}}
\newcommand{\cin}[1]{{?#1}}
\newcommand{\cout}[1]{{!#1}}
\newcommand{\parop}{\mathbin{|}}

\newcommand{\pnull}{\mathtt{0}}
\newcommand{\new}[1]{(\nu#1)}
\newcommand{\paction}{\ell}
\newcommand{\vin}[3]{#1?(#2:#3)}
\newcommand{\vout}[2]{#1!#2}
\newcommand{\chin}[2]{#1?(#2)}
\newcommand{\chout}[2]{#1!#2}
\newcommand{\bang}{{\star}}
\newcommand{\invisible}{\tau}

\newcommand{\fn}{\mathtt{fn}}
\newcommand{\bn}{\mathtt{bn}}
\newcommand{\dom}{\mathtt{dom}}

\newcommand{\subject}{\mathtt{subj}}

\newcommand{\win}{\checkmark}

\newcommand{\lred}[1]{\stackrel{#1}{\longrightarrow}}
\newcommand{\xlred}[1]{\xrightarrow{#1}}

\newcommand{\wlred}[1]{\stackrel{#1}{\Longrightarrow}}
\newcommand{\nwlred}[1]{\Longarrownot\wlred{#1}}

\newcommand{\nlred}[1]{\longarrownot\lred{#1}}

\newcommand{\rulename}[1]{{\textsc{(#1)}}}

\newcommand{\init}[1]{\init(#1)}

\newcommand{\subst}[2]{
	\{ 	\raisebox{.5ex}{\small$#1$}  /  
			\raisebox{-.5ex}{\small$#2$} \} 
	} 

\newcommand{\ssubc}{\sqsubseteq}
\newcommand{\subc}{\preceq}

\newcommand{\eqc}{\approx}
\newcommand{\seqc}{\simeq}

\newcommand{\eqdef}{\stackrel{\mathrm{def}}{=}}

\theoremstyle{definition}
\newtheorem{definition}{Definition}[section]
\newtheorem{example}{Example}[section]
\newtheorem{remark}{Remark}[section]

\theoremstyle{plain}
\newtheorem{proposition}{Proposition}[section]
\newtheorem{theorem}{Theorem}[section]

\newcommand{\marginnote}[2]{
  \ifcomments
  \marginpar{\parbox{2cm}{\flushleft \tiny \textbf{#1}: #2}}
  \fi
}
\newcommand{\Luca}[1]{\marginnote{Luca}{#1}}

\newcommand{\mytitle}{Session Types at the Mirror}

\title{\mytitle}

\author{
  Luca Padovani
  \institute{Istituto di Scienze e Tecnologie dell'Informazione,
    Universit\`a degli Studi di Urbino ``Carlo Bo''}
  \email{luca.padovani@uniurb.it}
}
\def\titlerunning{\mytitle}
\def\authorrunning{Luca Padovani}

\begin{document}
\maketitle

\begin{abstract}
  We (re)define session types as projections of process behaviors with
  respect to the communication channels they use. In this setting, we
  give session types a semantics based on fair testing. The outcome is
  a unified theory of behavioral types that shares common aspects with
  conversation types and that encompass features of both dyadic and
  multi-party session types.
  The point of view we provide sheds light on the nature of session
  types and gives us a chance to reason about them in a framework
  where every notion, from well-typedness to the subtyping relation
  between session types, is semantically -- rather than syntactically
  -- grounded.
\end{abstract}

\section{Introduction}
\label{sec:intro}

The \emph{leitmotif} in the flourishing literature on session
types~\cite{Honda1993,HondaVasconcelosKubo98,HondaYoshidaCarbone08} is
to associate every communication channel with a type that
\emph{constraints} how a process can use that channel. In this paper
we take the opposite perspective: we \emph{define} the session type
associated with a channel as the \emph{projection} of the behavior of
the processes restricted to how that channel is used by them. As
expected, this approach requires a language of session types that is
more general than the ones we usually encounter in other works. But --
this is in summary the contribution of this work -- the language we
come up with is just a minor variation of well-known value-passing
process algebras that can be semantically characterized using
well-known concepts and techniques.


To get acquainted with our approach, let us consider the following
example written in $\pi$-calculus like language and which is a
slightly simplified variant of the motivating example
in~\cite{HondaYoshidaCarbone08}:
\[
\begin{array}{@{}rcl@{}}
\text{Seller} & = & 
\chin{a}{x}.
\vin{x}{\mathit{title}}{\tstring}.
\vout{x}{\mathit{price}(\mathit{title})}.
\vin{x}{\mathit{addr}}{\taddress}.
\vout{x}{\mathit{date}(\mathit{title})}
\\
\text{Buyer1} & = &
\new{c}
\chout{a}{c}.
\vout{c}{\text{``The Origin of Species''}}.
\vin{c}{\mathit{price}}{\tint}.
\new{d}\chout{b}{d}.
\vout{d}{\mathit{price}/2}.
\chout{d}{c}
\\
\text{Buyer2} & = &
\chin{b}{y}.
\vin{y}{\mathit{contrib}}{\tint}.
\chin{y}{z}.
\vout{z}{\mathit{address}}.
\vin{z}{d}{\mathtt{Date}}
\end{array}
\]

Here we have two buyers that collaborate with each other in order to
complete a transaction with a seller. Buyer1 creates a local channel
$c$ that it sends to Seller through the public channel $a$. The
channel $c$ is normally dubbed \emph{session}: it is a fresh channel
shared by Buyer1 and Seller on which the two can communicate
privately. On $c$, Buyer1 sends to the Seller the name of a book, and
Seller answers with its price. At this stage Buyer1 asks for the
collaboration of Buyer2: it creates another fresh channel $d$ which it
communicates to Buyer2 by means of the public channel $b$, it sends
Buyer2 the amount of money Buyer2 should contribute, and finally it
\emph{delegates} the private channel $c$ to Buyer2, so that Buyer2 can
complete the transaction with the Seller. This implies sending the
Seller a delivery address and receiving the estimated delivery date.

Let us focus on the public channels $a$ and $b$: the former is used by
Buyer1 for sending a channel of some type, say $\contract$, and is
used by Seller for receiving a channel of the same type. In our
approach we say that the type of $a$ is $\cin{\contract}.\cwin \parop
\cout{\contract}.\cwin$, where $\cin{\contract}.\cwin$ is the
projected behavior of Seller on $a$, $\cout{\contract}.\cwin$ is the
projected behavior of Buyer1 on $a$, and $\parop$ denotes the
composition of these two behaviors.
In a similar way, $b$ is used by Buyer1 and Buyer2 and has type
$\cout{\varcontract}.\cwin \parop \cin{\varcontract}.\cwin$, assuming
that the channel exchanged between Buyer1 and Buyer2 has type
$\varcontract$.
Channel $c$ is more interesting: it is created by Buyer1, which uses
it according to the type $\cout{\tstring}.\cin{\tint}$. However, $c$
is delegated to Seller right after its creation, and to Buyer2 when
Buyer1 has finished using it. So, the true type of $c$ is
$\contract \parop \cout{\tstring}.\cin{\tint}.\client$ where
$\contract$ is the projection of Seller's behavior with respect to the
channel $c$ (after it has been received by Seller), and $\client$ is
the projection of Buyer2's behavior with respect to the same channel
after it has been received by Buyer2.
By similar arguments, one can see that the type of $d$ is
$\varcontract \parop \cout{\tint}.\cout{\client}.\cwin$ and the
mentioned types $\contract$, $\varcontract$, and $\client$ are defined
as $\cin{\tstring}.\cout{\tint}.\cin{\taddress}.\cout{\tdate}.\cwin$,
$\cin{\tint}.\cin{\client}.\cwin$, and
$\cout{\taddress}.\cin{\tdate}.\cwin$, respectively.
If we were to depict the projection we have operated for typing the
channels in the example, we could summarize it as follows:
\[
\begin{array}{rccccc}
  & \overbrace{\smash{\phantom{\cin{\tstring}.\cout{\tint}.\cin{\taddress}.\cout{\tdate}.\cwin}}}^\text{Seller}
  & & \overbrace{\smash{\phantom{\cout{\tstring}.\cin{\tint}.\cwin}}}^\text{Buyer1}
  & & \overbrace{\smash{\phantom{\cout{\taddress}.\cin{\tdate}.\cwin}}}^\text{Buyer2}
\\
  a: & \cin{\contract}.\cwin & & \cout{\contract}.\cwin \\
  b: & & & \cout{\varcontract}.\cwin & & \cin{\varcontract}.\cwin \\
  c: & \cin{\tstring}.\cout{\tint}.\cin{\taddress}.\cout{\tdate}.\cwin & & \cout{\tstring}.\cin{\tint}.\cwin & & \cout{\taddress}.\cin{\tdate}.\cwin \\
  d: & & & \cout{\tint}.\cout{\client}.\cwin & & \cin{\tint}.\cin{\client}.\cwin \\
\end{array}
\]

Can we tell whether the system composed of Seller and the two buyers
``behaves well''? Although at this stage we have not given a formal
semantics to session types, by looking at the types for the various
channels involved in the example we can argue that they all eventually
``reduce'' to a parallel composition of $\cwin$'s. If we read the type
$\cwin$ as the fact that a process stops using a channel with that
type, this roughly indicates that all the conversations initiated in
the example eventually terminate successfully.

The projection we have operated abstracts away from the temporal
dependencies between communications occurring on different
channels. This is a well-known source of problems if one is interested
in global progress properties.  In our approach, and unlike other
presentations of session types, we do not even try to impose any
linearity constraint on the channels being used, nor do we use
\emph{polarities}~\cite{GayHole05} or
indexes~\cite{HondaYoshidaCarbone08,BCDDDY08} for distinguishing
different \emph{roles}. For example, the process Buyer1 keeps using
channel $c$ after it has been delegated, and it delegates the channel
once more before terminating. As a consequence, the projection we
operate may not even capture the temporal dependencies between
communications occurring \emph{on the same channel}. This can happen
if two distinct free variables are instantiated with the same channel
during some execution. Thus, we must impose additional constraints on
processes only to ensure the type preservation
property. Interestingly, we will see that these additional constraints
are similar to those used for ensuring global
progress~\cite{DLY07,BCDDDY08,CastagnaDezaniGiachinoPadovani09}.

We can identify three main contributions of this work: (1) we show
that session types can be naturally generalized to an algebraic
language of processes that closely resembles value-passing {\ccs}; (2)
as a consequence, we are able to work on session types reusing a vast
toolkit of known results and techniques; in particular, we are able to
semantically justify the fundamental concepts (duality,
well-typedness, the subtyping relation) that are axiomatically or
syntactically presented in other theories; (3) we provide a unified
framework of behavioral types that encompasses features not only of
dyadic and multi-party session types, but also of conversation
types~\cite{CairesVieira09}.

\paragraph{Structure of the paper.}
In Section~\ref{sec:types} we define session types as a proper process
algebra equipped with a labeled transition system and a testing
semantics based on fair testing. This will immediately provide us with
a semantically justified equivalence relation -- actually, a pre-order
-- to reason about safe replacement of channels and well-behaving
systems.
In Section~\ref{sec:processes} we formally define a process language
that is a minor variant of the $\pi$-calculus without any explicit
construct that is dedicated to session-oriented interaction. We will
show how to type processes in this language and illustrate the main
features of the type system with several examples. Finally, we will
state the main properties (type preservation and local progress) of
our typing relation.
Section~\ref{sec:conclusions} concludes.

\paragraph{Related work.}
Theories of dyadic session types can be traced back to the works of
Honda~\cite{Honda1993} and Honda \emph{et
  al.}~\cite{HondaVasconcelosKubo98}. Since then, the application of
session types has been extended to functional
languages~\cite{VasconcelosGayRavara06,GayVasconcelos07} and
object-oriented languages (see \cite{DDC07,DDMY08} for just a few
examples). A major line of research is the one dealing with so-called
multi-party session types, those describing sessions where multiple
participants interact
simultaneously~\cite{HondaYoshidaCarbone08,BCDDDY08}.  An in depth
study of a subtyping relation for session types can be found
in~\cite{GayHole05}, while \cite{Vasconcelos09} provides an
incremental tutorial presentation of the most relevant features of
dyadic session types.

Conversation types~\cite{CairesVieira09} are a recently introduced
formalism that aims at generalizing session types for the description
of the behavior of processes that interact within and across the scope
of structurally organized communications called conversations.
Conversation types are very similar to the language of session types
we propose here, for example they embed a parallel composition
operator for representing the composed behavior of several processes
simultaneously accessing a conversation. The difference with our
approach mainly resides in the semantics of types: we treat session
types as terms of a proper process algebra with a proper transition
relation and all the relevant notions on types originate from
here. In~\cite{CairesVieira09}, the semantics of conversation types is
given in terms of syntactically-defined notions of subtyping and
\emph{merging}.\Luca{Controllare queste affermazioni.}
Also, \cite{CairesVieira09} uses a process language that incorporates
explicit constructs for dealing with conversations, while we emphasize
the idea of \emph{projected behavior} by working with the naked
$\pi$-calculus.

Elsewhere~\cite{CastagnaDezaniGiachinoPadovani09} we have been
advocating the use of a testing approach in order to semantically
justify session types. Unlike~\cite{CastagnaDezaniGiachinoPadovani09},
here we disallow branch selection depending on the type of
channels. This reduces the expressiveness of types for the sake of a
simplification of the technicalities in the resulting theory. Another
difference is that in the present paper we adopt a \emph{fair testing}
approach~\cite{RensinkVogler07}.

Finally, it should be mentioned that the use of processes as types has
already been proposed in the past, for example
in~\cite{TYPESMODELS,NIELSON}. In particular, \cite{NIELSON} uses a
language close to value-passing {\ccs} for defining an effect system
for Concurrent ML.



\section{Syntax and semantics of session types}
\label{sec:types}

Let us fix some conventions: $\contract$, $\varcontract$, $\client$,
\dots range over \emph{session types}; $\alpha$, \dots range over
\emph{actions}; $\type$, $\vartype$, \dots range over \emph{types};
$\val$, \dots range over an unspecified set $\mathcal{V}$ of
\emph{basic values}; $\tbasic$, \dots range over an unspecified set of
\emph{basic types} such as $\tint$, $\tbool$, $\tstring$, and so on.
The syntax of session types is defined by the grammar in
Table~\ref{tab:syntax}. Types represent sets of related values:
$\tempty$ is the empty type, the one inhabited by no value; basic
types are arbitrary subsets of $\mathcal{V}$; for every $\val \in
\mathcal{V}$ we write $\val$ for the \emph{singleton type} whose only
value is $\val$ itself.
%
We will write $\val : \type$ to state that $\val$ inhabits type
$\type$ and we will sometimes say that $\val$ is of type $\type$.

\begin{table}
\caption{\label{tab:syntax} Syntax of session types.}
\framebox[\textwidth]{
\begin{math}
\displaystyle
\begin{array}{@{}c@{\quad}c@{\quad}c@{}}
\begin{array}[t]{@{}rcl@{\quad}l@{}}
  \contract & ::= & & \text{\textbf{session type}} \\
  &   & \cnull & \text{(failure)} \\
  & | & \cwin & \text{(success)} \\
  & | & \alpha.\contract & \text{(action prefix)} \\
  & | & \contract + \contract & \text{(external choice)} \\
  & | & \contract \oplus \contract & \text{(internal choice)} \\
  & | & \contract \parop \contract & \text{(composition)} \\
\end{array}
&
\begin{array}[t]{@{}rcl@{\quad}l@{}}
  \alpha & ::= & & \text{\textbf{action}} \\
  &   & \cin{\type} & \text{(value input)} \\
  & | & \cout{\type} & \text{(value output)} \\
  & | & \cin{\contract} & \text{(channel input)} \\
  & | & \cout{\contract} & \text{(delegation)} \\
\end{array}
&
\begin{array}[t]{@{}rcl@{\quad}l@{}}
  \type & ::= & & \text{\textbf{type}} \\
  &   & \tempty & \text{(empty)} \\
  & | & \val & \text{(singleton)} \\
  & | & \tbasic & \text{(basic type)} \\
\end{array}
\end{array}
\end{math}
}
\end{table}

Actions represent input/output operations on a channel. An action
$\cout{\type}$ represents the sending of an arbitrary value of type
$\type$; an action $\cin{\type}$ represents the receiving of an
arbitrary value of type $\type$; actions $\cout{\contract}$ and
$\cin{\contract}$ are similar but they respectively represent the
sending and receiving of a channel of type $\contract$.

Although session types are used to classify channels, they describe
the behavior of processes using those channels. Consistently with this
observation, we will often present session types as characterizing
processes rather than channels. In the explanation that follows, it is
useful to keep in mind that, when a process uses a channel according
to some protocol described by a session type, it expects to interact
with other processes that use the same channel according to other
protocols. For a communication to occur, the process must perform an
action on the channel (say, sending a value of some type), and another
process must perform the corresponding co-action (say, receiving a
value of the same type).
The session type $\cnull$ classifies a channel on which a
communication error has occurred. No correct system should ever
involve channels typed by $\cnull$, and we will see that it is useful
to have an explicit term denoting a static error.
The session type $\cwin$ describes a process that performs no further
action on a channel.
The session type $\alpha.\contract$ describes a process that performs
the action $\alpha$, and then behaves according to the protocol
$\contract$.
The session type $\contract + \varcontract$ is the \emph{external
  choice} of $\contract$ and $\varcontract$ and describes a process
that offers interacting processes to behave according to one of the
branches.
Dually, the session type $\contract \oplus \varcontract$ is the
\emph{internal choice} of $\contract$ and $\varcontract$ and describes
a process that internally decides to behave according to one of the
branches.
The session type $\contract \parop \varcontract$ describes the
simultaneous access to a shared channel by two processes behaving
according to $\contract$ and $\varcontract$.\footnote{We use the word
  ``shared'' to highlight the fact that two (or more) processes
  simultaneously act on the same channel. This should not be confused
  with the terminology used in different session type theories, where
  ``shared channels'' are publicly known channels on which sessions
  are initiated.} If we have $n$ processes sharing a common channel
and each process behaves according to some protocol $\contract_i$,
then $\contract_1 \parop \cdots \parop \contract_n$ describes the
overall protocol implemented by the processes on the channel.

We do not rely on any explicit syntax for describing recursive
behaviors. We borrow the technique already used
in~\cite{CastagnaDezaniGiachinoPadovani09} and define the set of
session types as the set of possibly infinite syntax trees generated
by the productions of the grammar in Table~\ref{tab:syntax} that
satisfy the following conditions:
\begin{enumerate}
\item the tree must contain a finite number of \emph{different}
  subtrees;

\item on every infinite branch of the tree there must be infinite
  occurrences of the action prefix operator;

\item the tree must contain a finite number of occurrences of the
  parallel composition operator.
\end{enumerate}

The first condition is a standard \emph{regularity condition} imposing
that the tree must be a \emph{regular tree}~\cite{Courcelle83}. The
second one is a \emph{contractivity condition} ruling out meaningless
regular trees such as those generated by the equations $X = X + X$ or
$X = X \oplus X$. Finally, it can be shown that the last condition
enforces that the protocol described by a session type is ``finite
state''.

To familiarize with session types consider the following two examples:
\[\cin{\tint}.\cout{\tstring}.\cwin + \cin{\tbool}.\cout{\treal}.\cwin\]
describes a process that waits for either an integer number or a
Boolean value. If the process receives an integer number, it sends a
string; if the process receives a Boolean value, it sends a real
number. After that, in either case, the process stops using the
channel.
Instead, the session type $\cout{\tint}.\cwin \oplus
\cout{\tbool}.\cwin$ describes a process that internally decides
whether to send an integer or a Boolean value.

It may seem that the syntax of session types is overly generic, and
that external choices make sense only when they are guarded by input
actions and internal choices make sense only when they are guarded by
output actions. As a matter of facts, this is a common restriction in
standard session type presentations. In our approach, this generality
is actually \emph{necessary}: a session type $\contract =
\cout{\tint}.\cwin \parop \cout{\tbool}.\cwin$ describes two processes
trying to simultaneously send an integer and a Boolean value on the
same channel. A process interacting with these two parties is allowed
to read both values in either order, since both are available. In
other words, the session type $\contract$ is equivalent to
$\cout{\tint}.\cout{\tbool}.\cwin + \cout{\tbool}.\cout{\tint}.\cwin$,
that is the interleaving of the actions in $\contract$. Had we
expanded $\contract$ to $\cout{\tint}.\cout{\tbool}.\cwin \oplus
\cout{\tbool}.\cout{\tint}.\cwin$ instead, no interacting process
would be able to decide which value, the integer or the Boolean value,
to read first. The ability to express parallel composition in terms of
choices is well studied in process algebra communities where it goes
under the name of \emph{expansion
  law}~\cite{CCSWithoutTau,HennessyBook}. This ability is fundamental
in order to define complete proof systems and algorithms for deciding
equivalences. Decidability issues aside, we envision two more reasons
why this generality is appealing: first, it allows us to express the
typing rules (Section~\ref{sec:processes}) in a more compositional
way, which is particularly important in our approach where we aim at
capturing full, unconstrained process behaviors; second, it clearly
separates communications (represented by actions) from choices, thus
yielding a clean, algebraic type language with orthogonal features.

We equip session types with an operational semantics that mimics the
actions performed by processes behaving according to these types.  The
labeled transition system of session types is defined by the rules in
Table~\ref{tab:lts.sessions} plus the obvious symmetric rules of those
concerning choices and parallel composition. Transitions make use of
\emph{labels} ranged over by $\mu$, \dots and generated by the
grammar:
\[
\mu ~~::=~~ \win ~~\mid~~ \cin{\val} ~~\mid~~ \cout{\val} ~~\mid~~ \cin{\contract} ~~\mid~~ \cout{\contract}
\]

\begin{table}
\caption{\label{tab:lts.sessions} Transitions of session types.}
\framebox[\textwidth]{
\begin{math}
\displaystyle
\begin{array}{@{}c@{}}
\inferrule[\rulename{r1}]{}{
  \cwin \lred{\mathstrut\win} \cwin
}
\qquad
\inferrule[\rulename{r2}]{}{
  \contract \oplus \varcontract \lred{\mathstrut} \contract
}
\qquad
\inferrule[\rulename{r3}]{}{
  \cout{\val}.\contract \lred{\mathstrut\cout{\val}} \contract
}
\qquad
\inferrule[\rulename{r4}]{}{
  \cout{\client}.\contract \lred{\mathstrut\cout{\client}} \contract
}
\qquad
\inferrule[\rulename{r5}]{}{
  \cin{\client}.\contract \lred{\mathstrut\cin{\client}} \contract
}
\\\\
\inferrule[\rulename{r6}]{
  \val : \type
}{
  \cout{\type}.\contract \lred{\mathstrut} \cout{\val}.\type
}
\qquad
\inferrule[\rulename{r7}]{
  \val : \type
}{
  \cin{\type}.\contract \lred{\mathstrut\cin{\val}} \contract
}
\qquad
\inferrule[\rulename{r8}]{
  \contract \lred{\mathstrut} \contract'
}{
  \contract + \varcontract \lred{\mathstrut} \contract' + \varcontract
}
\qquad
\inferrule[\rulename{r9}]{
  \contract \lred{\mathstrut\mu} \contract'
}{
  \contract + \varcontract \lred{\mathstrut\mu} \contract'
}
\qquad
\inferrule[\rulename{r10}]{
  \contract \lred{\mathstrut} \contract'
}{
  \contract \parop \varcontract \lred{\mathstrut} \contract' \parop \varcontract
}
\\\\
\inferrule[\rulename{r11}]{
  \contract \lred{\mathstrut\mu} \contract'
  \\
  \mu \ne \win
}{
  \contract \parop \varcontract \lred{\mathstrut\mu} \contract' \parop \varcontract
}
\qquad
\inferrule[\rulename{r12}]{
  \contract \lred{\mathstrut\win} \contract'
  \\
  \varcontract \lred{\mathstrut\win} \varcontract'
}{
  \contract \parop \varcontract \lred{\mathstrut\win} \contract' \parop \varcontract'
}
\qquad
\inferrule[\rulename{r13}]{
  \contract \lred{\mathstrut\cout{\val}} \contract'
  \\
  \varcontract \lred{\mathstrut\cin{\val}} \varcontract'
}{
  \contract \parop \varcontract \lred{\mathstrut} \contract' \parop \varcontract'
}
\\\\
\inferrule[\rulename{r14}]{
  \contract \lred{\mathstrut\cout{\client}} \contract'
  \\
  \varcontract \lred{\mathstrut\cin{\client'}} \varcontract'
  \\
  \client \subc \client'
}{
  \contract \parop \varcontract \lred{\mathstrut} \contract' \parop \varcontract'
}
\qquad
\inferrule[\rulename{r15}]{
  \contract \lred{\mathstrut\cout{\client}} \contract'
  \\
  \varcontract \lred{\mathstrut\cin{\client'}} \varcontract'
  \\
  \client \not\subc \client'
}{
  \contract \parop \varcontract \lred{\mathstrut} \cnull
}
\end{array}
\end{math}
}
\end{table}

Strictly speaking, the transition system is defined by two relations:
a labeled one $\lred{\mu}$ describing \emph{external, visible actions}
and an unlabeled one $\lred{}$ describing \emph{internal, invisible
  actions}. Thus, the transition system is an extension of the one of
{\ccs} without $\tau$'s~\cite{CCSWithoutTau} to a value-passing
calculus.
Rule~\rulename{r1} states that the session type $\cwin$ emits a single
action $\win$ denoting successful termination of the protocol, and
reduces to itself.
By rule~\rulename{r2}, the session type $\contract \oplus
\varcontract$ can perform an internal transition to either $\contract$
or $\varcontract$.
Rules~\rulename{r3} and~\rulename{r4} deal with output actions. The
session type $\cout{\val}.\contract$ emits the value $\val$ and
reduces to $\contract$. Similarly, $\cout{\client}.\contract$ emits a
signal $\cout{\client}$ (the output of a channel of type $\client$).
Rule~\rulename{r5} is the dual of rule~\rulename{r4} and states that
$\cin{\client}.\contract$ emits a signal $\cin{\client}$ (the input of
a channel of type $\client$).
Rule~\rulename{r6} states that a process behaving according to
$\cout{\type}.\contract$ internally chooses a value $\val$ of type
$\type$ to send, and once has committed to such a value it reduces to
$\cout{\val}.\contract$.
Rule~\rulename{r7} is the dual of rule~\rulename{r3}, but because of
rule~\rulename{r6} observe that a process behaving according to
$\cout{\type}.\contract$ commits to sending \emph{one particular}
value of type $\type$, whereas a process behaving according to
$\cin{\type}.\contract$ is able to receive \emph{any} value of type
$\type$.
Rule~\rulename{r8} states that $+$ is indeed an external choice, thus
internal choices in either branch do not preempt the other
branch. This is a typical reduction rule for those languages with two
different choices, such as {\ccs} without
$\tau$'s~\cite{CCSWithoutTau}.
Rules~\rulename{r9} and~\rulename{r10} state obvious reductions for
external choices, which offer any action that is offered in either
branch, and parallel compositions, which allow either component to
internally evolve independently.
Rule~\rulename{r11} states that any action other than $\win$ is
offered by a parallel composition whenever it is offered by one of the
components; rule~\rulename{r12} states that a parallel composition has
successfully terminated only if both components have;
rule~\rulename{r13} states the obvious synchronization between
components offering dual actions.
Rule~\rulename{r14} states that a process sending a channel of type
$\client$ can synchronize with another process willing to receive a
channel of type $\client'$, but only if $\client \subc \client'$. Here
$\subc$ is a \emph{subtyping relation} meaning that any channel of
type $\client$ can be used where a channel of type $\client'$ is
expected. We shall formally define $\subc$ in a moment; for the time
being we must content ourselves with this intuition.
Rule~\rulename{r15} states that if the relation $\client \subc
\client'$ is not satisfied, the synchronization occurs nonetheless,
but it yields an error.

Before we move on to the subtyping relation for session types, we
should point out a fundamental design decision that relates
communication and external choices.
On the one hand, values other than channels may drive the selection of
the branch in external choices. For example, we have
$\cin{\tint}.\contract + \cin{\tbool}.\varcontract \lred{\cin{3}}
\contract$ while $\cin{\tint}.\contract + \cin{\tbool}.\varcontract
\xlred{\cin{\ttrue}} \varcontract$. The \emph{type} of the value
determines the branch, and this feature allows us to model the
label-driven branch selection that is found in standard
session types theories.
On the other hand, the last two rules in Table~\ref{tab:lts.sessions}
show that branch selection cannot be affected by the type of the
channel being communicated. It is true that $\cin{\client}.\contract +
\cin{\client'}.\varcontract \lred{\cin{\client}} \contract$ and
$\cin{\client}.\contract + \cin{\client'}.\varcontract
\lred{\cin{\client'}} \varcontract$, but when we compose
$\cin{\client}.\contract + \cin{\client'}.\varcontract$ with
$\cout{\client''}.\varcontract'$ either reduction is possible, and the
residual may or may not be $\cnull$ depending on the relation between
$\client$, $\client'$, and $\client''$:
\[
\inferrule{
  \client'' \subc \client
}{
  \cin{\client}.\contract + \cin{\client'}.\varcontract \parop \cout{\client''}.\varcontract' \lred{} \contract \parop \varcontract'
}
\qquad\qquad\qquad
\inferrule{
  \client'' \not\subc \client
}{
  \cin{\client}.\contract + \cin{\client'}.\varcontract \parop \cout{\client''}.\varcontract' \lred{} \cnull
}
\]

To be sure that the residual is not $\cnull$, it must be the case that
$\client'' \subc \client$ \emph{and} $\client'' \subc \client'$.
In summary, we do not allow dynamic dispatching according to the type
of a channel, namely all channels are treated as if they had the same
type. This is not the only possible choice
(see~\cite{CastagnaDezaniGiachinoPadovani09} for an alternative), but
is one that simplifies the theory.

In the following we adopt standard conventions regarding the
transition relations: we write $\wlred{}$ for the reflexive,
transitive closure of $\lred{}$; we write $\contract \lred{\mu}$
(respectively, $\contract \wlred{\mu}$) if there exists $\varcontract$
such that $\contract \lred{\mu} \varcontract$ (respectively,
$\contract \wlred{\mu} \varcontract$); we write $\nlred{}$,
$\nlred{\mu}$, $\nwlred{\mu}$ for the usual negated relations; for
example, $\contract \nlred{}$ means that $\contract$ does not perform
internal transitions.

The first semantic characterization we give is that of \emph{complete}
session type, namely a session type that can always reach a successful
state, no matter of its internal transitions.

\begin{definition}[completeness]
\label{def:completeness}
  We say that $\contract$ is \emph{complete} if $\contract \wlred{}
  \contract'$ implies $\contract' \wlred{\win}$.
\end{definition}

Intuitively, a complete protocol is one implemented by processes which
can always terminate successfully their interaction, without the help
of any other process. Observe, as a side note, that completeness
implies that no evolution of the system may yield an error or lead to
a state where one process insists on sending a message that no
interacting party is willing to accept.
$\cwin$ is the simplest complete session type; the session types
$\cin{\contract}.\cwin \parop \cout{\contract}.\cwin$ and
$\contract \parop \cout{\tstring}.\cin{\tint}.\client$ we have seen in
the introduction are also complete, since every maximal transition
leads to a successfully terminated state.
The simplest example of incomplete session type is $\cnull$, another
example being $\cin{\tint}.\cwin \parop \cout{\treal}.\cwin$ because
of the maximal reduction $\cin{\tint}.\cwin \parop \cout{\treal}.\cwin
\lred{} \cin{\tint}.\cwin \parop \cout{\sqrt{2}}.\cwin \nlred{}$.
If we take $\contract$ as the solution of the equation $X =
\cin{\tint}.X$ and $\varcontract$ as the solution of the equation $Y =
\cout{\tint}.Y$ we have that $\contract \parop \varcontract$ is
\emph{not} complete, despite it never reaches a deadlock state. In
this sense the notion of completeness embeds a \emph{fairness
  principle} that is typically found in fair testing
theories~\cite{RensinkVogler07}.

Completeness is the one notion that drives the rest of the theory. We
define the subtyping relation for session types, which we call
\emph{subsession}, as the relation that preserves completeness:
$\contract$ is ``smaller than'' $\varcontract$ if every session type
that completes $\contract$ completes $\varcontract$ as well.

\begin{definition}[subsession]
\label{def:subs}
We say that $\contract$ is a \emph{subsession} of $\varcontract$,
notation $\contract \subc \varcontract$, if $\contract \parop \client$
complete implies $\varcontract \parop \client$ complete for every
$\client$.
  We write $\eqc$ for the equivalence relation induced by $\subc$,
  namely ${\eqc} = {\subc} \cap {\succeq}$.
\end{definition}

In other words, we are defining an equivalence relation for session
types based on (fair) testing~\cite{RensinkVogler07}: we use
completeness as the notion of test, and we say that two session types
are equivalent if they pass the same tests. As a consequence, the
equational theory generated by this definition is not immediately
obvious, although a few relations are easy to check: for example, $+$,
$\oplus$, and $\parop$ are commutative, associative operators;
$\cnull$ is neutral for $+$ and $\cwin$ is neutral for $\parop$;
furthermore $\contract \oplus \varcontract \subc \contract$. Namely,
it is safe to use a channel with type $\contract \oplus \varcontract$
where another one of type $\contract$ is expected.
If the safety property mentioned here seems hard to grasp, one should
resort to the intuition that the ``type'' of a channel actually is the
behavior of a process communicating on that channel. A process that
expects to receive a channel with type $\contract$ will behave on that
channel according to $\contract$; if we send that process a channel
with type $\contract \oplus \varcontract$, the receiving process will
still behave according to $\contract$, which is a more deterministic
behavior than $\contract \oplus \varcontract$, hence no problem may
arise.
As a special case of reduction of nondeterminism, we have
$\cout{\treal}.\contract \subc \cout{\tint}.\contract$ assuming that
$\tint$ is a subtype of $\treal$.
Other useful relations are those concerning failed processes: we have
$\cnull \eqc \alpha.\cnull$ and $\cout{\tempty}.\contract \eqc
\cin{\tempty}.\contract \eqc \cnull$. More generally, the relation
$\contract \eqc \cnull$ means that there is no session type
$\varcontract$ such that $\contract \parop \varcontract$ is complete:
$\contract$ is intrinsically flawed and cannot be remedied. The class
of non-flawed session types will be of primary importance in the
following, to the point that we reserve them a name.

\begin{definition}[viability]
\label{def:viability}
We say that $\contract$ is \emph{viable} if $\contract \parop \client$
is complete for some $\client$.
\end{definition}

\begin{remark}
  At this stage we can appreciate the fact that subsession depends on
  the transition relation, and that the transition relation depends on
  subsession. This circularity can be broken by stratifying the
  definitions: a session type $\contract$ is given weight $0$ if it
  contains no prefix of the form $\cin{\client}$ or $\cout{\client}$;
  a session type $\contract$ is given weight $n > 0$ if any session
  type $\client$ in any prefix of the form $\cin{\client}$ or
  $\cout{\client}$ occurring in $\contract$ has weight at most $n -
  1$. By means of this stratification, one can see that the
  definitions of the transition relation and of subsession are well
  founded.
  \eoe
\end{remark}

It is fairly easy to see that $\subc$ is a precongruence with respect
to action prefix, internal choice, and parallel composition.
The case of the action prefix is trivial.
As regards the internal choice, it suffices to observe that
$(\contract \oplus \varcontract) \parop \client$ is complete if and
only if both $\contract \parop \client$ and $\varcontract \parop
\client$ are complete. Namely, $\oplus$ corresponds to a set-theoretic
intersection between session types that complete $\contract$ and
$\varcontract$.
As regards the parallel composition, the precongruence follows from
the very definition of subsession, since $\contract \parop \contract'
\subc \varcontract \parop \contract'$ if $(\contract \parop
\contract') \parop \client$ complete implies $(\varcontract \parop
\contract') \parop \client$ complete, namely if $\contract \parop
(\contract' \parop \client)$ complete implies $\varcontract \parop
(\contract' \parop \client)$ complete, that is if $\contract \subc
\varcontract$.
Because all the non-viable session types are $\eqc$-equal, however,
$\subc$ is \emph{not} a precongruence with respect to the external
choice. For example, we have $\cnull \subc \cout{\tint}.\cnull$ but
$\cout{\tint}.\cwin + \cnull \not\subc \cout{\tint}.\cwin +
\cout{\tint}.\cnull \eqc \cnull$.
This is a major drawback of the subsession relation as it is defined,
since it prevents $\subc$ from being used in arbitrary contexts for
replacing equals with equals (note that $\eqc$ is \emph{not} a
congruence for the same reasons).
We resort to a standard technique for defining the largest relation
included in~$\subc$ that is a precongruence with respect to the
external choice. We call this relation \emph{strong subsession}:

\begin{definition}[strong subsession]
\label{def:ssubc}
  Let $\ssubc$ be the largest relation included in $\subc$ that is a
  precongruence with respect to $+$, namely $\contract \ssubc
  \varcontract$ if and only if $\contract + \client \subc \varcontract
  + \client$ for every $\client$.
  We write $\seqc$ for the equivalence relation induced by $\ssubc$,
  namely ${\seqc} = {\ssubc} \cap {\sqsupseteq}$.
\end{definition}

We end this section with a few results about $\subc$ and
$\ssubc$. First of all, we can use $\ssubc$ for reasoning about
viability and completeness of a session type:

\begin{proposition}
\label{prop:ssubc}
  The following properties hold:
\begin{enumerate}
\item $\contract$ is not viable if and only if $\contract \ssubc
  \cnull$;

\item $\contract$ is complete if and only if $\cwin + \contract \ssubc
  \contract$.
\end{enumerate}
\end{proposition}

Then, we prove that $\subc$ and $\ssubc$ are \emph{almost} the same
relation, in the sense that they coincide as soon as the smaller
session type is viable. This means that for all practical purposes the
use of $\ssubc$ in place of $\subc$ is immaterial, if not for the
gained precongruence, since in no case we will be keen on replacing a
channel with a viable type with one that is not viable.

\begin{theorem}
\label{thm:strong.subc}
  $\contract \subc \varcontract$ if and only if either $\contract
  \ssubc \cnull$ or $\contract \ssubc \varcontract$.
\end{theorem}

\begin{remark}
  It is interesting to compare $\ssubc$ with the subtyping relation
  for session types in~\cite{GayHole05}.  From a technical point of
  view, the two relations arise in completely different ways: $\ssubc$
  arises semantically as a relation between session types that
  preserves completeness; the subtyping relation in~\cite{GayHole05}
  is defined (co)inductively and by cases on the syntax of session
  types being related.
  The essence of this latter relation is strictly connected with the
  direction of the exchanged messages: when $S \leq T$ holds, $S$
  sends more things and receives fewer, regardless of whether such
  things are labels or actual data. In contrast the relation $\ssubc$
  is fundamentally determined by reduction of nondeterminism, which is
  captured by the law $\contract \oplus \varcontract \ssubc
  \contract$. Note that this law does not say anything about messages
  being sent or received. For example, we have $\cout{\tint}.\cwin
  \oplus \cout{\tbool}.\cwin \ssubc \cout{\tint}.\cwin$ but also
  $\cin{\tint}.\cwin \oplus \cin{\tbool}.\cwin \ssubc
  \cin{\tint}.\cwin$.
  We can identify two other significant differences: the first one is
  that in our theory of session types, the successfully terminated
  session type $\cwin$ can be composed with actions. For example,
  $\cwin + \cin{\tint}.\cwin$ describes a process that is waiting for
  an integer, but is also perfectly happy to terminate the session at
  this time without any further communication. As another example,
  $\cwin \oplus \cout{\tint}.\cwin$ describes the behavior of a
  process that internally decides whether to terminate the session
  without any further communication, or to do so only after having
  sent an integer. Incidentally, observe that the two examples
  complete each other. In~\cite{GayHole05} (and in most session type
  theories) the terminal behavior cannot be composed with others. The
  type system we will describe later does not use this capability
  either, but this is just to keep it simple and with a reasonable
  number of rules.
  The second and last difference we want to emphasize is that the law
  $\contract \ssubc \contract + \varcontract$, which is somehow dual
  of $\contract \oplus \varcontract \ssubc \contract$, does \emph{not}
  hold, while it is sound in~\cite{GayHole05}. Two main reasons
  justify this fact: the first is that in our theory $+$ is an
  algebraic operator that can combine arbitrary session types, and for
  this reason the external choice sometimes is an internal choice in
  disguise: for example, it is possible to prove that
  $\cin{\tint}.\contract + \cin{\tint}.\varcontract \seqc
  \cin{\tint}.(\contract \oplus \varcontract)$. This cannot happen
  in~\cite{GayHole05} because of the very syntax of session types,
  which prevents arbitrary compositions of behaviors. The second reason
  is that the synchronous communication model we are relying upon does
  not tolerate the introduction of \emph{interferences}. For example,
  we have $\cin{\tint}.\cwin \not\ssubc \cin{\tint}.\cwin +
  \cin{\tbool}.\cwin$ because the session type $\cout{\tint}.\cwin +
  \cout{\tbool}.\cnull$ completes the first session type but not the
  second one: the $\cin{\tbool}.\cnull$ branch introduces an
  interference that may enable harmful synchronizations. For this and
  other reasons the adoption of a synchronous communication model is
  questionable in practice. However, one can show that by suitably
  restricting behaviors (for instance, by forbidding outputs in
  external choices such as in the example above) some instances of the
  law $\contract \ssubc \contract + \varcontract$ become sound
  again. Furthermore, it is possible to simulate partial forms of
  asynchrony by means of the session type language we have presented
  (the idea is not explored in detail here, but the interested reader
  may find some hints in~\cite{CastagnaPadovani09}).
  In summary, the $\ssubc$ relation is both an extension and a
  conservative restriction of the subtyping relation
  in~\cite{GayHole05}.
\eoe
\end{remark}



\section{Processes}
\label{sec:processes}

\begin{table}
\caption{\label{tab:syntax.process} Syntax of processes.}
\framebox[\textwidth]{
\begin{math}
\displaystyle
\begin{array}{@{}c@{\qquad}c@{}}
\begin{array}[t]{@{}rcll@{}}
  \process & ::= & & \text{\textbf{process}} \\
  & | & \pnull & \text{(idle)} \\
  & | & \pi.\process & \text{(action prefix)} \\
  & | & \bang\process & \text{(replication)} \\
  & | & \process + \process & \text{(external choice)} \\
  & | & \process \oplus \process & \text{(internal choice)} \\
  & | & \process \parop \process & \text{(parallel composition)} \\
  & | & \new{c}\process & \text{(restriction)} \\
\end{array}
& 
\begin{array}[t]{@{}rcll@{}}
  \pi & ::= & & \text{\textbf{action}} \\
  & | & \vin{u}{x}{\type} & \text{(value input)} \\
  & | & \vout{u}{\expr} & \text{(value output)} \\
  & | & \chin{u}{x} & \text{(channel input)} \\
  & | & \chout{u}{v} & \text{(delegation)} \\
\end{array}
\end{array}
\end{math}
}
\end{table}

Processes are defined by the grammar in
Table~\ref{tab:syntax.process}. We use $\process$, $\varprocess$,
$\clientp$, \dots to range over processes; we use $\pi$, \dots to
range over action prefixes; we use $a$, $b$, $c$, \dots to range over
\emph{channel names}; we let $x$, $y$, $z$, \dots range over
\emph{variables} and $u$, $v$, \dots range over channel names and
variables ($v$ should not be confused with $\val$ that we used to
range over elements of $\mathcal{V}$); we let $e$, \dots range over an
unspecified language of \emph{expressions}.
The process language is a minor variation of the $\pi$-calculus, so we
remark here only the differences:
we have four action prefixes: $\vin{u}{x}{\type}$ denotes a receive
action for a basic value $x$ of type $\type$ on channel $u$;
$\vout{u}{\expr}$ denotes a send action for the value of the
expression $\expr$ on channel $u$; $\chin{u}{x}$ denotes a receive
action for a channel $x$ on channel $u$; $\chout{u}{v}$ denotes a send
action for a channel $v$ on channel $u$. Consistently with the
language of session types, actions denoting send/receive operations of
channels are ``untyped''.
The process $\bang\process$ denotes unbounded replications of process
$\process$, and $\process + \varprocess$ and $\process \oplus
\varprocess$ respectively denote the external and internal choice
between $\process$ and $\varprocess$.
We will usually omit the $\pnull$ process; we will write
$\fn(\process)$ for the set of free channel names occurring in
$\process$ (the only binder for channel names is restriction); we will
write $\process\subst{m}{x}$ for the process $\process$ where all free
occurrences of the variable $x$ have been replaced by~$m$.

\begin{table}
\caption{\label{tab:lts.processes} Transitions of processes.}
\framebox[\textwidth]{
\begin{math}
\displaystyle
\begin{array}{@{}c@{}}
\inferrule{}{
  \process \oplus \varprocess \lred{\mathstrut\invisible} \process
}
\qquad
\inferrule{}{
  \bang\process \lred{\mathstrut\invisible} \bang\process \parop \process
}
\qquad
\inferrule{}{
  c?(x).\process \lred{\mathstrut c\ain{d}} \process\subst{d}{x}
}
\qquad
\inferrule{}{
  c!d.\process \lred{\mathstrut c\aout{d}} \process
}
\\\\
\inferrule{
  \val : \type
}{
  \vin{c}{x}{\type}.\process \lred{\mathstrut c\cin{\val}} \process\subst{\val}{x}
}
\qquad
\inferrule{
  \expr \downarrow \val
}{
  \vout{c}{e}.\process \lred{\mathstrut c\cout{\val}} \process
}
\qquad
\inferrule{
  \process \lred{\mathstrut\invisible} \process'
}{
  \process + \varprocess \lred{\mathstrut\invisible} \process' + \varprocess
}
\qquad
\inferrule{
  \process \lred{\mathstrut\paction} \process'
  \\
  \paction \ne \invisible
}{
  \process + \varprocess \lred{\mathstrut\paction} \process'
}
\\\\
\inferrule{
  \process \lred{\mathstrut c\aout{m}} \process'
  \\
  \varprocess \lred{\mathstrut c\ain{m}} \varprocess'
}{
  \process \parop \varprocess \lred{\mathstrut\invisible} \process' \parop \varprocess'
}
\quad
\inferrule{
  \process \lred{\mathstrut c\aout{(d)}} \process'
  \\
  \varprocess \lred{\mathstrut c\ain{d}} \varprocess'
  \\
  d \not\in \fn(\varprocess)
}{
  \process \parop \varprocess \lred{\mathstrut\invisible} \new{d}(\process' \parop \varprocess')
}
\quad
\inferrule{
  \process \lred{\mathstrut\paction} \process'
  \\
  \bn(\paction) \cap \fn(\varprocess) = \emptyset
}{
  \process \parop \varprocess \lred{\mathstrut\paction} \process' \parop \varprocess
}
\\\\
\inferrule{
  \process \lred{\mathstrut\paction} \process'
  \\
  d \not\in \fn(\paction) \cup \bn(\paction)
}{
  \new{d}\process \lred{\mathstrut\paction} \new{d}\process'
}
\qquad
\inferrule{
  \process \lred{\mathstrut c\aout{d}} \process'
  \\
  c \ne d
}{
  \new{d}\process \lred{\mathstrut c\aout{(d)}} \process'
}
\end{array}
\end{math}
}
\end{table}

The transition relation of processes is defined by an almost standard
relation in Table~\ref{tab:lts.processes}, so we will not provide
detailed comments here. In the table, we write $\expr \downarrow \val$
for the fact that expression $\expr$ evaluates to $\val$.
Labels of the transition relation are ranged over by $\paction$, \dots
and are generated by the following grammar:
\[
\paction ~~::=~~ \invisible ~~\mid~~ c\cin{m} ~~\mid~~ c\cout{m} ~~\mid~~ c\cout(d)
\]
where $m$, \dots ranges over \emph{messages}, namely basic values and
channel names.
Action~$\invisible$ denotes an internal computation or a
synchronization.
Actions of the form $c\cin{m}$ and $c\cout{m}$ are often called
\emph{free inputs} and \emph{free outputs} respectively.
Actions of the form $c\cout(d)$ are called \emph{bound outputs} and
represent the extrusion of a private channel, $d$ in this case. We use
these actions to model session initiations, whereby a private channel
is exchanged and subsequently used for the actual interaction.
Notions of free and bound names in labels are standard, with
$\fn(c\cin{d}) = \fn(c\cout{d}) = \{ c, d \}$, $\fn(c\cin{\val}) =
\fn(c\cout{\val}) = \fn(c\cout(d)) = \{ c \}$, and $\bn(c\cout(d)) =
\{ d \}$, the other sets being empty.

We remark only two distinctive features of the transition relation:
(1) the replicated process $\bang\process$ evolves by means of an
internal transition to $\bang\process \parop \process$; technically
this makes $\bang\process$ a \emph{divergent process}, but the fact
that we work with a fair semantics makes this only a detail; (2)
similarly to the transition relation for session types, the transition
relation for processes selects branches of external choices according
to the type of the basic value being communicated. This is evident in
the transitions for $\vin{c}{x}{\type}.\process$, which are labeled
by values of type $\type$.

\begin{table}
\caption{\label{tab:typing} Typing rules for processes.}
\framebox[\textwidth]{
\begin{math}
\begin{array}{@{}c@{}}
\inferrule[\rulename{t-weak}]{
  \Gamma \vdash \process : \Delta
  \\
  u \not\in \dom(\Delta)
}{
  \Gamma \vdash \process : \Delta \cup \{ u : \cwin \}
}
\qquad
\inferrule[\rulename{t-sub}]{
  \Gamma \vdash \process : \Delta \cup \{ u : \varcontract \}
  \\
  \contract \ssubc \varcontract
}{
  \Gamma \vdash \process : \Delta \cup \{ u : \contract \}
}
\qquad
\inferrule[\rulename{t-res}]{
  \Gamma \vdash \process : \Delta \cup \{ c : \cwin + \contract \}
}{
  \Gamma \vdash \new{c}\process : \Delta
}
\\\\
\inferrule[\rulename{t-nil}]{\ }{
  \Gamma \vdash \pnull : \emptyset
}
\qquad
\inferrule[\rulename{t-input}]{
  \Gamma, x : \type \vdash \process : \Delta \cup \{ u : \contract \}
}{
  \Gamma \vdash \vin{u}{x}{\type}.\process : \Delta \cup \{ u : \cin{\type}.\contract \}
}
\qquad
\inferrule[\rulename{t-inputS}]{
  \Gamma \vdash \process : \{ x : \client \}
}{
  \Gamma \vdash \chin{u}{x}.\process : \{ u : \cin{\client}.\cwin \}
}
\\\\
\inferrule[\rulename{t-output}]{
  \Gamma \vdash \expr : \type
  \\
  \Gamma \vdash \process : \Delta \cup \{ u : \contract \}
}{
  \Gamma \vdash \vout{u}{\expr}.\process : \Delta \cup \{ u : \cout{\type}.\contract \}
}
\qquad
\inferrule[\rulename{t-outputS}]{
  \Gamma \vdash \process : \Delta \cup \{ u : \contract, v : \varcontract \}
}{
  \Gamma \vdash \chout{u}{v}.\process : \Delta \cup \{ u : \cout{\client}.\contract, v : \varcontract \parop \client \}
}
\\\\
\inferrule[\rulename{t-ext}]{
  \Gamma \vdash \pi_i.\process_i : \Delta \cup \{ u : \contract_i \}~{}^{i\in I}
  \\
  \subject(\pi_i) = u~{}^{i\in I}
}{
  \Gamma \vdash \sum_{i\in I} \pi_i.\process_i : \Delta \cup \{ u : \sum_{i\in I} \contract_i \}
}
\qquad
\inferrule[\rulename{t-int}]{
  \Gamma \vdash \process : \Delta
  \\
  \Gamma \vdash \varprocess : \Delta
}{
  \Gamma \vdash \process \oplus \varprocess : \Delta
}
\\\\
\inferrule[\rulename{t-bang}]{
  \Gamma \vdash \process : \{ u_i : \contract_i~{}^{i\in I} \}
  \\
  \contract_i \ssubc (\contract_i \parop \contract_i)~{}^{i\in I}
}{
  \Gamma \vdash \bang\process : \{ u_i : \contract_i~{}^{i\in I} \}
}
\qquad
\inferrule[\rulename{t-par}]{
  \Gamma \vdash \process : \{ u_i : \contract_i~{}^{i\in I} \}
  \\
  \Gamma \vdash \varprocess : \{ u_i : \varcontract_i~{}^{i\in I} \}
}{
  \Gamma \vdash \process \parop \varprocess : \{ u_i : \contract_i \parop \varcontract_i~{}^{i\in I} \}
}
\end{array}
\end{math}
}
\end{table}

The typing rules for the process language are inductively defined in
Table~\ref{tab:typing}. Judgments have the form $\Gamma \vdash
\process : \Delta$, where $\Gamma$ is a standard environment mapping
variables to basic types and $\Delta$ is an environment mapping
channel names and channel variables to session types. We write
$\dom(\Delta)$ for the domain of~$\Delta$.
Rule~\rulename{t-weak} allows one to enrich $\Delta$ with assumptions
of the form $u : \cwin$, indicating that a process does not use the
channel $u$. The premise $u \not\in \dom(\Delta)$ implies $u \not\in
\fn(\process)$ since it is always the case that $\fn(\process)
\subseteq \dom(\Delta)$.
Rule~\rulename{t-sub} is an almost standard subsumption rule regarding
the type of a channel $u$. The peculiarity is that it works ``the
other way round'' by allowing a session type $\varcontract$ to become
a smaller session type $\contract$. The intuition is that $\process$
behaves according to $\varcontract$ on the channel $u$. Thus, it is
safe to declare that the session type associated with $u$ is even less
deterministic than $\varcontract$. This rule is fundamental in the
type system since many other rules impose equality constraints on
session types that can only be satisfied by finding a lower bound to
two or more session types. It should also be appreciated the
importance of using $\ssubc$, which is a precongruence, since this
allows us to apply rule~\rulename{t-sub} in arbitrary contexts.
Rule~\rulename{t-res} types restrictions, by requiring the session
type associated with the restricted channel to be of the form $\cwin +
\contract$. In light of rule~\rulename{t-sub} and of
Proposition~\ref{prop:ssubc}(2), this requirement imposes that the
session type of a restricted channel $c$ must be complete. Namely,
there must not be communication errors on $c$.
Rule~\rulename{t-nil} types the idle process $\pnull$ with the empty
session environment. The process $\pnull$ should not be confused with
the failed session type $\cnull$: the former is the successfully
terminated process that does not use any channel; the latter denotes a
communication error or a deadlock.
Rule~\rulename{t-input} types an input action for basic values of type
$\type$. The assumption $x : \type$ is moved into the environment
$\Gamma$ and if the session type associated with $u$ in the
continuation $\process$ is $\contract$, then the overall behavior of
$\process$ on $u$ is described by $\cin{\type}.\contract$.
Rule~\rulename{t-output} is similar, but regards output actions of
basic values. We assume an unspecified set of deduction rules for
judgments of the form $\Gamma \vdash \expr : \type$, denoting that the
expression $\expr$ has type $\type$ in the environment $\Gamma$.
Rule~\rulename{t-inputS} types an input action for a channel $x$. The
continuation $\process$ must be typed in a session environment of the
form $\{ x : \client \}$, requiring that $\process$ must not refer to
(free) channels other than the received one. Consequently, the whole
process behaves according to the session type
$\cin{\client}.\cwin$. The severe restriction on the continuation
process is necessary for type preservation, as we will see in
Example~\ref{ex:inputs} below.
Rule~\rulename{t-outputS} types delegations, whereby a channel $v$ is
sent over another channel $u$. This rule expresses clearly the idea of
projection we are pursuing in our approach: the delegated channel $v$
is used in the continuation $\process$ according to the session type
$\varcontract$ (which may be $\cwin$ in case rule~\rulename{t-weak} is
applied); at the same time, the channel $v$ is delegated to another
process which will behave on it according to $\client$. As a
consequence, the overall behavior on $v$ is expressed by the
composition of $\varcontract$ and $\client$, namely by
$\varcontract \parop \client$.  If $u$ is used in the continuation
$\process$ according to $\contract$, then its type is
$\cout{\client}.\contract$ in the conclusion.
Rule~\rulename{t-ext} types external choices. These are well typed
only when each branch of the choice is guarded by an action whose
subject is $u$ (we write $\subject(\pi)$ for the subject of action
$\pi$). For this reason the rule is only applicable to processes of
the form $\pi_1.\process_1 + \cdots + \pi_n.\process_n$, which we
abbreviate as $\sum_{i\in\{1,\dots,n\}} \pi_i.\process_i$, and the
resulting behavior on $u$ is the sum $\contract_1 + \cdots +
\contract_n$ of the individual behaviors on $u$ of each branch, which
we abbreviate as $\sum_{i\in I} \contract_i$.
Rule~\rulename{t-int} types internal choices, but only when the two
branches do have the same session environment. This can be achieved by
repeated applications of rules~\rulename{t-weak} and~\rulename{t-sub}.
Rule~\rulename{t-par} types the parallel composition of
processes. Again this rule shows the idea of projection and, unlike
other session type systems, allows (actually requires) both processes
to use exactly the same channels, whose corresponding session types
are composed with $\parop$. In this context rule~\rulename{t-weak} can
be used to enforce that the session environments for $\process$ and
$\varprocess$ are exactly the same, recalling that $\cwin$ is neutral
for $\parop$.
Finally, rule~\rulename{t-bang} types replicated processes: the basic
idea is that a replicated process $\bang\process$ is well typed if any
channel it uses is ``unlimited'' (in the terminology
of~\cite{GayHole05}), which in our case translates to the property
that it must be smaller than two copies of itself. $\cwin$ is the
simplest session type with this property, but there are others as we
will see in Example~\ref{ex:bang}.

\begin{remark}
  Thanks to our setting, we have the opportunity to make some
  interesting connections between the subtyping relations used in type
  theories for programming languages and the behavioral preorders that
  arise in many testing theories for process algebras. According to
  Definition~\ref{def:subs}, if $\contract \subc \varcontract$, then
  it is safe to replace a process behaving according to~$\contract$
  with another process behaving according to~$\varcontract$. This is
  because, by definition of $\subc$, every context that completes
  $\contract$ will also complete $\varcontract$. Note in particular
  that the safe substitution regards the \emph{larger object}. This
  contrasts with the subtyping relations where it is safe to replace
  an object of type $T$ with another object of type $S$ if $S$ is a
  subtype of $T$. In fact, this is exactly the notion of safe
  substitutability we are using in rule~\rulename{t-sub}. This
  mismatch can be source of confusion: recall that in our view a
  session type is not the type of channel, but rather is the allowed
  behavior of a process on a channel. Thus, if a channel has type
  $\cout{\tint}.\cwin$, that means that the process using it behaves
  according to $\cout{\tint}.\cwin$. Now, it is safe to replace that
  channel with another one with type $\cout{\tint}.\cwin \oplus
  \cout{\tbool}.\cwin$: since we are replacing the channel, and not
  the process, the process will still behave according to
  $\cout{\tint}.\cwin$, but this time on a channel that allows more
  behaviors. Since $\cout{\tint}.\cwin \oplus \cout{\tbool}.\cwin
  \ssubc \cout{\tint}.\cwin$, we are assured that the substitution is
  safe.
  \eoe
\end{remark}

\begin{example}[persistent service provider]
\label{ex:bang}
Consider the process
\[
\varprocess \equiv \bang\chin{\mathit{server}}{x}.\process
\]
which accepts an unbounded number of connection requests on the
channel $\mathit{server}$ and processes them in the process
$\process$. Assume we can type the non-replicated process as follows:
\[
\inferrule{
  \Gamma \vdash \process : \{ x : \client \}
}{
  \Gamma \vdash \chin{\mathit{server}}{x}.\process : \{ \mathit{server} : \cin{\client}.\cwin \}
}
\]

To apply rule~\rulename{t-bang} for $\varprocess$ we need
$\mathit{server}$ to have a type $\contract$ such that $\contract
\ssubc \contract \parop \contract$, and $\cin{\client}.\cwin$ clearly
does not have this property.
Consider the session type $\contract$ that is solution of the equation
$X = \cwin \oplus \cin{\client}.X$. We have $\contract \ssubc
\cin{\client}.\cwin$ and furthermore $\contract \ssubc
\contract \parop \contract$. Hence we can now type $\varprocess$ with
an application of rule~\rulename{t-sub} followed by~\rulename{t-bang}.
\eoe
\end{example}

\begin{example}[multi-party session]
  Intuitively, a multi-party session is a conversation taking place on
  a restricted channel that is shared between three or more
  participants.
  Consider a system $\new{a}(\process \parop \process \parop
  \varprocess)$ where
\[
\begin{array}{rcl}
  \process & \eqdef & \chin{a}{x}.\vin{x}{y}{\tint}.(\vout{x}{\mathit{isprime}(y)} + \vin{x}{z}{\mathtt{abort}}) \\
  \varprocess & \eqdef & \new{c}(\chout{a}{c}.\vout{c}{n} \parop \chout{a}{c}.\vout{c}{n} \parop \vin{c}{x}{\tbool}.\vout{c}{\mathtt{abort}}) \\
\end{array}
\]
the idea being that the two instances of $\process$ represent two
servers checking whether a number is prime. The process $\varprocess$
establishes a connection by sending the two servers a fresh channel
$c$ and sending on this channel some integer number $n$. The two
servers are thus able to process the number in parallel and the first
one that succeeds sends the result back to $\varprocess$. Upon
reception of the result from one of the servers, $\varprocess$
notifies the other server by sending a dummy value $\mathtt{abort}$,
which we assume is a singleton type inhabited only by $\mathtt{abort}$
itself.

It is easy to verify that, within $\process$, the channel $x$ has type
$\contract = \cin{\tint}.(\cout{\tbool}.\cwin +
\cin{\mathtt{abort}}.\cwin)$ and $a$ is used according to the type
$\cin{\contract}.\cwin$. In $\varprocess$, $a$ is used according to
the type $\cout{\contract}.\cwin \parop \cout{\contract}.\cwin$ and
$c$ is used according to the type $\contract \parop
\cout{\tint}.\cwin \parop \contract \parop \cout{\tint}.\cwin \parop
\cin{\tbool}.\cout{\mathtt{abort}}.\cwin$. Hence, the overall type of
$a$ is $\cout{\contract}.\cwin \parop \cout{\contract}.\cwin \parop
\cin{\contract}.\cwin \parop \cin{\contract}.\cwin$ and the whole
system is well typed since both $a$'s type and $c$'s type are
complete.
\eoe
\end{example}

The type system permits to find type derivations for processes using
channels with a non-viable session type. Examples of such processes
are $\vin{c}{x}{\tempty}.\pnull$. A non-viable session type indicates
an intrinsic flaw in the process. For this reason viability is really
the one notion that characterizes \emph{well-typedness} of processes.
We say that a session environment $\Delta$ is viable if so is every
session type in its codomain.

\begin{theorem}[subject reduction]
\label{thm:sr}
Let $\Gamma \vdash \process : \Delta$ and $\process \lred{\invisible}
\varprocess$ and $\Delta$ viable. Then $\Gamma \vdash \varprocess :
\Delta$.
\end{theorem}

\begin{example}
  If compared with more standard session type theories, the notion of
  \emph{viability} looks as an additional complication of our more
  general setting. Actually, the rules in Table~\ref{tab:typing}
  project the behavior of a process with respect to the channels it
  uses and impose a few local constraints. Then, the viability
  hypothesis in Theorem~\ref{thm:sr} ensures that the process is
  really well behaved. Without this hypothesis, subject reduction does
  not hold. Consider for example the process $\process \parop
  \varprocess$ where
\[
  \process \eqdef \new{c}(\chout{a}{c})
  \qquad
  \varprocess \eqdef \chin{a}{x}.\vout{x}{3}
\]

On one hand, $\process$ sends a fresh channel $c$ to $\varprocess$,
but does not use it anymore. On the other hand, $\varprocess$ pretends
to send an integer on the channel it receives from
$\process$. According to the rules in Table~\ref{tab:typing} we have
$\vdash \process \parop \varprocess : \{ a :
\cout{(\cwin)}.\cwin \parop \cin{(\aout{\tint}.\cwin)}.\cwin \}$. In
particular, the session type associated with $a$ is not viable,
because $\cwin \subc \aout{\tint}.\cwin$ does not hold. Indeed, we
have the reduction
\[
  \process \parop \varprocess
  \lred{\invisible}
  \new{c}(\pnull \parop \vout{c}{3})
\]
where the residual process is ill typed, since $c$ is associated with
the session type $\cwin \parop \aout{\tint}.\cwin$ which is not
complete, hence it does not satisfy the premise of
rule~\rulename{t-res}.
\eoe
\end{example}

Before addressing type safety, we justify by means of examples the two
main constraints imposed by the type system in order to guarantee type
preservation.

\begin{example}
\label{ex:ext}
To justify rule~\rulename{t-ext}, consider the process
\[
  \process \eqdef \vin{a}{x}{\tint}.\vin{b}{y}{\tbool} + \vin{b}{x}{\tint}.\vin{a}{y}{\tbool}
\]
and suppose it well typed, where $a : \cin{\tint}.\cwin +
\cin{\tbool}$ and $b : \cin{\tbool}.\cwin + \cin{\tint}.\cwin$.
Apparently, \emph{both} $a$ and $b$ are able to receive either an
integer or a Boolean value and a system such as
$\new{a}\new{b}(\process \parop \vout{a}{3} \parop \vout{b}{3})$ would
be well typed. Alas, the external choices in the types of $a$ and $b$
do not take into account the fact that any synchronization of
$\process$ with another process may actually \emph{disable} one branch
in these choices.
The reduction
\[
\new{a}\new{b}(\process \parop \vout{a}{3} \parop \vout{b}{3})
\lred{\invisible}
\new{a}\new{b}(\vin{b}{y}{\tbool} \parop \pnull \parop \vout{b}{3})
\]
leads to an ill-typed process, since $b$ has type
$\cin{\tbool}.\cwin \parop \cout{\tint}.\cwin$ which is not complete.
\eoe
\end{example}

\begin{example}
\label{ex:inputs}
The severe constraint in the premise of rule~\rulename{t-inputS} can
be justified by looking at the following processes:
\[
\begin{array}{rcl}
  \process & \eqdef & \chout{a}{c}.\chout{a}{c}.\vin{c}{x}{\tint}.\vin{c}{y}{\tbool} \\
  \varprocess & \eqdef & \chin{a}{x}.\chin{a}{y}.\vout{y}{\ttrue}.\vout{x}{3} \\
\end{array}
\]
where $\process$ can be typed with a derivation like the following:
\[
\inferrule{
  \inferrule{
    \inferrule{
      \vdots
    }{
      \Gamma \vdash
      \vin{c}{x}{\tint}.\vin{c}{y}{\tbool} 
      : \{ a : \cwin, c : \cin{\tint}.\cin{\tbool}.\cwin \}
    }
  }{
    \Gamma \vdash
    \chout{a}{c}.\vin{c}{x}{\tint}.\vin{c}{y}{\tbool} 
    : \{ a : \cout{(\cout{\tbool}.\cwin)}.\cwin, c : \cout{\tbool}.\cwin \parop \cin{\tint}.\cin{\tbool}.\cwin \}
  }
}{
  \Gamma \vdash
  \chout{a}{c}.\chout{a}{c}.\vin{c}{x}{\tint}.\vin{c}{y}{\tbool} 
  : \{ a : \cout{(\cout{\tint}.\cwin)}.\cout{(\cout{\tbool}.\cwin)}.\cwin,
       c : \cout{\tint}.\cwin \parop \cout{\tbool}.\cwin \parop \cin{\tint}.\cin{\tbool}.\cwin
  \}
}
\]

The process $\process$ delegates the channel $c$ twice on $a$. The
first time, the delegated behavior is $\cout{\tint}.\cwin$, while the
second time the delegated behavior is $\cout{\tbool}.\cwin$. Each time
$c$ is delegated, $\process$ assumes that the receiving process will
implement the delegated behavior. However, as it can be clearly seen
in the conclusion of the typing derivation above, the overall
delegated behavior of $c$ is $\cout{\tint}.\cwin \parop
\cout{\tbool}.\cwin$, namely the parallel composition of the two
behaviors that were separately delegated. This is fundamental for the
completeness of $c$'s type, since the input operations performed by
the residual of $\process$ at the top of the typing derivation occur
in a specific order.

The process $\varprocess$, which receives both delegations, is unaware
that $x$ and $y$ will be instantiated with the same channel. So,
$\varprocess$ is well typed and $x$ and $y$ have respectively type
$\cout{\tbool}.\cwin$ and $\cout{\tint}.\cwin$, as requested by
$\process$, but $\varprocess$ uses these channels in a specific order
that is not captured by the projections.  The process $\process \parop
\varprocess$ deadlocks in two steps:
\[
  \process \parop \varprocess
  \lred{\invisible}
  \lred{\invisible}
  \vin{c}{x}{\tint}.\vin{c}{y}{\tbool} \parop \vout{c}{\ttrue}.\vout{c}{3}
\]
where in the final state we have $c :
\cin{\tint}.\cin{\tbool}.\cwin \parop
\cout{\tbool}.\cout{\tint}.\cwin$ which is not complete.
By requiring, in the premise of rule~\rulename{t-inputS}, that the
receiving process cannot use any channel other than the received one,
we are basically imposing that the receiving process must handle every
received channel in a thread of its own. 
\eoe
\end{example}

In judgments of the form $\Gamma \vdash \process : \Delta$ the
environment $\Delta$ is an approximation of $\process$ insofar as it
describes the projections of $\process$'s behavior with respect to the
channels it uses and delegates. It is well known that this
approximation is unable to capture situations where well-typed
processes deadlock because the interdependence between communications
occurring on different channels are lost. Our approach is no
exception, as shown by the following example.

\begin{example}[deadlock]
  Consider the system
\[
  \new{a}\new{b}(\vout{a}{3}.\vin{b}{x}{\tbool} \parop \vout{b}{\ttrue}.\vin{a}{x}{\tint})
\]
where the channels $a$ and $b$ have respectively type $\contract =
\cout{\tint}.\cwin \parop \cin{\tint}.\cwin$ and $\varcontract =
\cin{\tbool}.\cwin \parop \cout{\tbool}.\cwin$.
In both cases we have $\cwin + \contract \ssubc \contract$ and $\cwin
+ \varcontract \ssubc \varcontract$, hence the system is well typed
but deadlock.
\eoe
\end{example}

The safety property we are able to state guarantees that, if all the
processes sharing some channel $c$ are immediately ready to
communicate on $c$, then they will eventually synchronize. Since in
our transition relation for processes synchronization is triggered not
just by the channels on which messages are exchanged, but also by the
type of the exchanged messages, the eventual synchronization
translates to the fact that there is no communication error: it is
never the case that there is a process willing to send a message of
some type, and no other process is ever willing to receive messages of
that particular type.  The notion of ``readiness'' we mentioned is
captured by the following definition:

\begin{definition}[readiness]
  We say that $\process$ is \emph{ready} on $c$ if $\process
  \downarrow c$ is derivable by the rules:
\[
\begin{array}{@{}c@{}}
  \inferrule{}{
    \pi.\process \downarrow \subject(\pi)
  }
  \qquad
  \inferrule{
    c \not\in \fn(\process)
  }{
    \process \downarrow c
  }
  \qquad
  \inferrule{
    \process \downarrow c
    \\
    \varprocess \downarrow c
  }{
    \process + \varprocess \downarrow c
  }
  \qquad
  \inferrule{
    \process \downarrow c
    \\
    \varprocess \downarrow c
  }{
    \process \parop \varprocess \downarrow c
  }
  \qquad
  \inferrule{
    \process \downarrow c
    \\
    c \ne d
  }{
    \new{d}\process \downarrow c
  }
\end{array}
\]
\end{definition}

Intuitively, $\process$ is ready on $c$ if either it does not use $c$,
in which case it plays no role in any synchronization on $c$, or if
$\process$ is prefixed by an action whose subject is $c$, or if every
branch of $\process$ is ready on $c$.
Observe that when $\process \equiv \process_1 + \process_2$,
\emph{both} branches are required to be ready on $c$. This is not
overly restrictive because, by rule~\rulename{t-ext}, if either branch
is prefixed by an action whose subject is $c$, so must be the other
branch.

\begin{theorem}
\label{thm:progress}
  If $\Gamma \vdash \process : \Delta \cup \{ c : \contract \}$ and
  $\contract$ complete and $\process \downarrow c$, then either $c
  \not\in \fn(\process)$ or $\process \lred{\invisible}$.
\end{theorem}


\section{Concluding remarks}
\label{sec:conclusions}

It may sound obvious to state that session types are behavioral
types. Yet, although session types are normally associated with
channels, channels do not expose any behavior. The solution of this
apparently innocuous paradox lays in the equally obvious observation
that the session type associated with a channel \emph{reflects} the
behavior of a \emph{process} concerning the input/output operations
that the process performs on that channel. By taking this mirrored
point of view we have been able to define a simple and, in our
opinion, elegant theory of session types that generalizes, unifies,
and semantically justifies many concepts that can be found scattered
in the current literature: (multi-party) session types are terms of a
suitably defined process algebra closely based on value-passing
{\ccs}; \emph{completeness} expresses the property that a session is
well-formed and never yields a communication error;
\emph{duality}~\cite{GayHole05} $\contract \bowtie \varcontract$ is
the special case where $\contract \parop \varcontract$ is complete;
\emph{viability} captures the concept of well-typed process, namely of
process that can be composed with others in order to implement
complete sessions; the \emph{subtyping relation} between session types
arises semantically by relating those session types that preserve
completeness in arbitrary contexts.

The adoption of a fair testing semantics~\cite{RensinkVogler07} for
session types is original to the best of our knowledge. In fact, most
presentations of session types rely on notions of duality or
well-formed composition where the only concern is the absence of
communication errors, while the fairness principle we adopt imposes an
additional constraint: that at any time a conversation is always able
to reach a so-called successful state. Whether or not this is
desirable in practice, from a technical point of view there are both
pros and cons: on the one hand, the fair subsession relation is more
difficult to characterize coinductively and axiomatically because
fairness escapes the mere structure of types; on the other hand, the
subsession relation is an all-in-one tool that incorporates safe
substitutability (rule~\rulename{t-sub}), viability, and completeness
(Proposition~\ref{prop:ssubc}). We have been unable to fully
characterize completeness in terms of a non-fair subsession relation
(see~\cite{CastagnaPadovani09} for an attempt in the context of
behavioral contracts).


The type system we have provided as a proof-of-concept in
Section~\ref{sec:processes} may look excessively restrictive, in
particular with respect to the rule~\rulename{t-inputS} which demands
that the continuation cannot use any (known) session if not the
received one. We have three observations regarding this point: (1)
this is a direct consequence of our focus on the idea of projected
behavior, which allows a more liberal use of channels; (2) similar
restrictions can be found in type systems guaranteeing global
progress~\cite{DLY07,BCDDDY08,CastagnaDezaniGiachinoPadovani09}; (3)
the provided type system is very natural and simple, considering the
freedom it leaves in the use of channels; this simplicity suggests
that it can be smoothly extended with features such as polarities or
roles which would likely help relaxing the constraints. We leave this
extension as future work.



\bibliographystyle{eptcs}
\bibliography{main}

\paragraph{Acknowledgments.} I wish to thank Giuseppe Castagna and
Mariangiola Dezani for having provided comments on early versions of
this paper. The anonymous referees of the ICE workshop have
contributed with invaluable feedback and insight, not only with their
reviews but also on the forum associated with the workshop Web site,
where they asked several intriguing questions.


\end{document}